\documentclass{article}

\usepackage[T1]{fontenc}

\usepackage{epsfig}

\usepackage{amsmath,amssymb,latexsym}
\usepackage{graphicx}

\newtheorem{theorem}{Theorem}[section]
\newtheorem{proposition}[theorem]{Proposition}

\newtheorem{lemma}[theorem]{Lemma}

\def\e{{\mathrm e}}
\def\C{{\mathbb C}}
\def\eff{{\cal F}}

\def\I{{\mathrm i}}
\renewcommand{\i}{{\mathrm i}}

\def\N{{\mathbb N}}

\def\R{{\mathbb R}}

\def\e{\mathrm{e}}
\def\d{\mathrm{d}}
\def\H{{\mathcal H}}

\newcommand{\ux}{{\underline x}}
\newcommand{\uk}{{\underline k}}

\def\Hd{{\mathcal H}_{\mathrm D}}
\def\A{{\mathcal A}}
\def\Ad{{\mathcal A}_{\mathrm D}}
\def\Af{{\mathcal A}_{\mathrm F}}
\def\Ar{{\mathcal A}_{\mathrm F, R}}
\def\Al{{\mathcal A}_{\mathrm F, L}}

\newcommand{\bbbone}{\mathchoice {\rm 1\mskip-4mu l} {\rm 1\mskip-4mu l}
{\rm 1\mskip-4.5mu l} {\rm 1\mskip-5mu l}}

\renewcommand{\tilde}{\widetilde}
\newcommand{\dom}{{\rm Dom}}

\begin{document}
\title{The Unruh effect revisited}
\author{S. De Bi\`evre\footnote{Stephan.De-Bievre@math.univ-lille1.fr}\\Universit\'e des Sciences et Technologies de Lille\\UFR de
Math\'ematiques et Laboratoire Painlev\'e\\59655 Villeneuve
d'Ascq, \ France
\and
M. Merkli\footnote{merkli@math.mcgill.ca; current address: Deptartment of
  Mathematics, University of Toronto, Toronto, Ontario, Canada M5S 2E4}\\Department of Mathematics and
Statistics\\
McGill University\\
805 Sherbrooke St. W.\\
Montr\'eal H3A 2K6, \ Canada}

\maketitle

\begin{abstract} We give a complete and rigorous proof of the Unruh effect, in the following form.
We show that the state of a two-level system, uniformly accelerated with
proper acceleration $a$, and coupled to a scalar bose  field initially in the
Minkowski vacuum state will converge, asymptotically in the detector's proper
time, to the Gibbs state at inverse temperature $\beta=\frac{2\pi}{a}$. The
result also holds if the field and detector are initially in an excited state.
We treat the problem as one of return to equilibrium, exploiting in particular
that the Minkowski vacuum is a KMS state with respect to Lorentz boosts. We
then use the recently developed spectral techniques to prove the stated
result.
\end{abstract}

\section{Introduction}
The following observation, now referred to as the Unruh effect, was made by W.
Unruh in 1976 \cite{u}. When a detector, coupled to a
 relativistic quantum field in its vacuum
 state,  is uniformly accelerated through Minkowski
 spacetime, with proper acceleration $a$, it registers a {\em  thermal} black body
radiation at temperature
 $T=\frac{\hbar a}{2\pi c k_{\mathrm B}}$. This is the so-called Unruh
 temperature.  In more anthropomorphic terms \cite{uw}, ``for a free quantum
 field in its vacuum state in Minkowski spacetime $M$ an observer with uniform
 acceleration $a$ will feel that he is bathed by a thermal distribution of
 quanta of the field at temperature $T$.''  This result has attracted a fair amount of
 attention, and generated considerable surprise and even some scepticism. For a review of
 various aspects of the subject, nice physical discussions of the phenomenon,
and further references, we refer to \cite{ta}, \cite{wa} \cite{fuu}. The
reason for the surprise is that, if you think of the vacuum as ``empty
space'', then you will find it puzzling that a detector, accelerated or not,
which may itself initially be in its ground state, will  ``see  particles'',
since, after all, in the vacuum, there  aren't any. In order not to be
surprised, one has to remember that, of course, the vacuum is not ``empty
space'', but
 the ground state of the field, and one should {\em expect}   the detector
to react to the presence of the field when it is accelerated through space.

For example, if you were to drag a detector along a non-relativistic chain of
oscillators in its ground state, you would certainly expect the coupling
between the detector and the oscillators to excite both.  The energy for this
process is, in final analysis, furnished by the agent that drags the detector
along the oscillator chain.

What  is nevertheless still surprising in connection with the Unruh effect is
the claim that the detector  ``perceives'' a {\em thermal} distribution of
radiation at some particular temperature that only depends on the
acceleration. To see what is precisely meant by these statements, it is
helpful to get rid of the anthropomorphic terminology used above and in much
of the literature as well as of all reference to particles or quanta, which
turn out to be irrelevant to the discussion. This is what we will do below. It
is worth pointing out in this connection that already in \cite{u},
``detection of a particle'' is defined by ``excitation of the detector'', and
does therefore not presuppose the actual definition of what a particle
precisely is, which is a tricky thing to do, as is well known \cite{fu}. In
fact, the computations in the physics literature of the excitation probability
of the detector can be seen to be  perturbative computations of the asymptotic
state of the detector (see \cite{uw} for example).  We therefore adopt the
following simple formulation of the Unruh effect. Consider the coupled
detector-field system.  Suppose that initially  it is in a  product
state with the field in the vacuum state. Now let the coupled system evolve.
At some later (detector proper) time, the state of the system will no longer
be a product state. Now trace out the field variables, to obtain the reduced
state of the detector (which will be a mixed state, even if the initial state
was pure). The Unruh effect states that the latter converges,
 asymptotically in the observer's proper time, to the Gibbs state at
the aforementioned temperature $T$. Note that this is not by any means
obvious: after all, a priori, it is not clear why the detector state should,
asymptotically in time, converge at all, and even if it does, it is not
obvious it should tend to a positive temperature state: a priori,  it could
have been any other mixed state.

It is our goal in this paper to give a complete and rigorous proof of the
above statement. The way we have formulated it makes it clear already that we
think of it as a problem in the theory of open quantum systems in which a
small system, here the detector, is coupled to a reservoir, here the field.
Let us formulate our result somewhat more precisely. For a completely rigorous
statement, we refer to Section \ref{s:model}. The model we consider is the one
proposed in \cite{uw}, which is itself a simplification of the model
considered in \cite{u}. The detector is modeled by a two-level system and the
field is taken to be a massive or massless Klein-Gordon field.  The observable
algebra of the detector  is therefore generated  by ``fermionic''
creation/annihilation operators $A$, $A^\dagger(\sigma)$. The free Heisenberg
evolution of the detector is 
$\dot A(\sigma)=-i E A(\sigma)$, where $\sigma$ is the detector's proper
time. In other words, the free detector Hamiltonian is 
$$
H_{\mathrm D}=E A^\dagger A.
$$
The coupling between the field and the detector
is realized via a monopole, and is ultraviolet regularized; it is sometimes referred to as a de
Witt monopole detector (see \cite{ta}). Suppose initially the detector-field
system is in a product state $\omega_0$ with the detector in a state described
by some density matrix $\rho$ and the field in the Minkowski vacuum state. Let
$B$ be a detector observable and $\alpha_\sigma^\lambda(B)$ its Heisenberg
evolution under the coupled dynamics, with coupling constant $\lambda$. Then
we prove that
\begin{equation}\label{eq:unruh1}
\lim_{\sigma\to\infty}\omega_0(\alpha_\sigma^\lambda(B))
=\frac{1}{Z_{\beta,\mathrm D}} \mathrm{Tr}\ \e^{-\beta
  H_{\mathrm D}}B+ O(\lambda^2).
\end{equation}
Here $\beta=(k_{\mathrm B}T)^{-1}$ with $T$ the Unruh temperature and $Z_{\beta, \mathrm
D}=\mathrm{Tr}\e^{-\beta H_{\mathrm D}}$. The approach to the equilibrium
state is exponentially fast. 

Our proof of this result is based on techniques developed in the last decade to
prove ``return to equilibrium'' in open quantum systems \cite {JP1,JP2,BFS,M,DJ,DJP}. We combine these with the
Bisognano-Wichman theorem \cite{biwi}, which states that the vacuum is a KMS
state for the Lorentz boosts on the Rindler wedge. The relevance of this last
result to the Unruh effect (and a generalization to more general spacetimes)
was explained a long time ago by Sewell in \cite{se}. Let us point out that
the work of Sewell, together with known stability results of KMS states
(see e.g. \cite{Da,KFGV}) imply a result somewhat similar to but considerably weaker than
(\ref{eq:unruh1}), namely
\begin{equation}\label{eq:unruh2}
\lim_{\sigma\to\infty,
  \lambda\to0,\lambda\sigma^2=1}\omega_0(\alpha_\sigma^\lambda(B))
=\frac{1}{Z_{\beta,\mathrm D}} \mathrm{Tr}\ \e^{-\beta
  H_{\mathrm D}}B.
\end{equation}
This is the so-called van Hove weak coupling limit. 
In our result, the
limit $\sigma\to\infty$ is shown to exist for all sufficiently small $\lambda$,
and to coincide with the right hand side of (\ref{eq:unruh1}).

The paper is organized as follows. In Section \ref{s:model}, we describe the
model in detail and state our main result. We will also comment on the precise
role played by the choice of the form factor determining the ultraviolet cutoff
in the interaction term. Section \ref{s:proof} is devoted to
its proof. The latter uses Araki's perturbation theory for KMS states and its
recent extensions, together with the spectral approach to the problem of
return to equilibrium developed in the cited references. Since this material
is rather technical, we have made an effort to state the result in Section
\ref{s:model} with as little reference to it as possible.\\

\noindent{\bf Acknowledgements:} SDB gratefully acknowledges the hospitality of McGill University,
Universit\'e de Montr\'eal and Concordia University, where part of this
work was performed. MM is grateful to Universit\'e des Sciences et
Technologies de Lille 1 for support and hospitality.


\section{The model and the result} \label{s:model}
We need to give a precise description of the model and in particular of its
dynamics. This requires some preliminaries.

\subsection{The free field}

Let us start by describing in detail the field to which the detector will be
coupled. The field operators are represented on the symmetric Fock space
$\eff$ over $L^2(\R^d, \d \underline x)$. Here $d\geq 1$ is the dimension of
space and  $x=(x^0,\ux)$ be is a point in Minkowski space--time $\R\times\R^d$
(with metric signature $(+,-,\dots,-)$). So $\eff:=\oplus_{n\in\N} \eff^{(n)}$,
where $\eff^{(n)}$ is the $n$-fold symmetric tensor product of the
one-particle space $L^2(\R^d, \d \underline x)$.

Let ${\cal S}(\R^{d+1};\R)$ and ${\cal S}(\R^{d+1};\C)$ denote the real  and
the complex valued Schwartz functions on $\R^{d+1}$, respectively. For
$f\in{\cal S}(\R^{d+1};\C)$, one defines the field operators in the usual way:
$$
 \Omega=(-\Delta+m^2)^{1/2}, S^{\pm}f=\int_\R \d t \frac1{\sqrt\Omega}\e^{\pm\I\Omega t}f_t, \
Q[f]=\frac1{\sqrt2}(a^\dagger(S^+ f) + a(\overline{S^-f})).
$$
Here $\Delta$ is the Laplacian, $m\geq0$ the mass, and $a, a^\dagger$ are the
usual creation and annihilation operators on $\eff$ (we follow the convention
that $f\mapsto a^\dagger(f)$ is linear while $f\mapsto a(f)$ is antilinear),
and the bar denotes complex conjugation. When $m=0$, we will
suppose $d>1$. Writing formally
$$
Q[f]=\int_{\R^{d+1}}\d x f(x) Q(x), \quad x=(x^0, \ux),
$$
this leads to the familiar
\begin{equation}
Q(x)=\int_{\R^d}\frac{\d\uk}{\sqrt{2\omega(\uk)}}\big[ \e^{\i \uk\,\ux
-\i\omega(\uk) x^0}a(\uk) +\e^{-\i \uk\,\ux +\i\omega(\uk) x^0}a^*(\uk)\big],
\label{b2}
\end{equation}
where $\omega(\uk)=\sqrt{\underline k^2 + m^2}$. The field satsifies the
Klein-Gordon equation $\square Q(x) + m^2 Q(x)=0$, where
$\square=\partial_{x^0}^2-\Delta$. We use units in which $\hbar=1=c$.

As can be learned in any book on special relativity (such as \cite{ri}), in an
adapted choice of inertial coordinate frame,  a uniformly accelerated
worldline of proper acceleration $a>0$, parametrized by its proper time
$\sigma$, has the form
$$
x^0(\sigma)=\frac1a\sinh a\sigma,\quad x^1(\sigma)=\frac1a\cosh a\sigma,\quad
x^2(\sigma)=0= x^3(\sigma).
$$
Associated to this worldline is the right wedge (or Rindler wedge) $
W_{\mathrm R}:=\{x\in\R^4 |\  x^1>|x^0|\}. $ It is the intersection of the
causal future and past of the worldline, or the collection of spacetime points
to which the observer on the worldline can send signals and from which he can
also receive signals. Note  for later reference that the left wedge
$W_{\mathrm L}:=-W_{\mathrm R}$ is the causal complement of $\overline
W_{\mathrm R}$.

There exists a global coordinate system  on $W_{\mathrm R}$ that is
particularly well adapted to the description of the problem at hand. It is
given by the so-called Rindler coordinates $(\tau,u,\underline
x_\perp)\in\R\times \R_+^*\times \R^{d-1}$, defined by
\begin{equation}\label{eq:coordchange}
x^0= u\sinh \tau,\ x^1=u\cosh \tau,\ \underline x_\perp=(x_2, \dots x_d).
\end{equation}
Here $\tau$ is a global time coordinate on the right wedge. Note that, given
$a\in\R^+, (\alpha_2,\dots \alpha_d)\in\R^{d-1}$, the curve $u=1/a$,
 $\underline x_\perp = (\alpha_2, \dots, \alpha_d)$ is the worldline of a
uniformly accelerated observer with proper acceleration $a$ and proper time
$\sigma = a^{-1} \tau$. In addition, two points in the right wedge with the
same value for the $\tau$-coordinate are considered as simultaneous in the
instantaneous rest frame of any such observer (see \cite{ri}).  Among the
Lorentz boosts, only the boosts in the $x_1$-direction leave the right wedge
invariant. In inertial coordinates they are given by the linear
transformations
$$
B_{\tau'}= \left[
\begin{array}{ccccc}
\cosh\tau' &\sinh\tau'&0&\dots &0\\
\sinh\tau'&\cosh\tau'&0&\dots &0\\
0&0&1&\dots&0\\
\vdots&\vdots&\vdots&\ddots&\vdots\\
0&0&0&\dots&1
\end{array}
\right].
$$
In the Rindler coordinates, this becomes $B_{\tau'}(\tau, u, \underline
x_\perp)= (\tau +\tau', u, \underline x_\perp)$. In this sense, the boosts in the
$x_1$-direction act as time translations on the Rindler wedge.

Since the field satisfies the Klein-Gordon equation, one has, in Rindler
coordinates on $W_{\mathrm R}$:
\begin{equation}\label{eq:kleingordonwedge}
\left(u^{-2}\partial_\tau^2 - u^{-1}\partial_{u} u
\partial_{u}  +(-\Delta_{\perp}+ m^2)\right) Q(\tau, u,\underline
x_\perp)=0.
\end{equation}
Moreover, the covariance of the free field under the Poincar\'e group yields,
for all $\tau\in\R$,
\begin{equation}\label{eq:boostfield}
Q[f\circ B_{-\tau}]=\e^{\I L_{\mathrm F} \tau}Q[f]\e^{-\I
  L_{\mathrm F} \tau},
\end{equation}
with
\begin{equation}\label{eq:liouvfreefield}
L_{\mathrm F}=\d \Gamma (K), \ K=\Omega^{1/2} X^1\Omega^{1/2},
\end{equation}
where $X^1$ is the operator of multiplication by $x^1$. 
In particular, for $x=(\tau, u, \underline x_\perp)\in W_{\mathrm R}$,
$$
Q(\tau, u, \underline x_\perp) = \e^{\I L_{\mathrm F}
  \tau}Q(0,u,\underline x_\perp)\e^{-\I  L_{\mathrm F} \tau}.
$$
In other words, $L_{\mathrm F}$ generates the free Heisenberg dynamics of the
field operators associated to the right wedge. Let us furthermore introduce,
for later purposes, the conjugate field
\begin{equation}\label{eq:conjfield}
P[f]:=\frac{\d }{\d \tau} Q[f\circ B_{-\tau}]\mid_{\tau=0}\,=\I\left[L_{\mathrm
F}, Q[f]\right].
\end{equation}
It then follows from the basic properties of the free field that the equal
time commutation relations of the field and the conjugate field are, at
$\tau=0$,
\begin{equation}\label{eq:equaltime}
\left[ Q(0, u, \underline x_\perp), P(0, u', \underline x_\perp')\right] =\I u
\delta_u(u')\ \delta_{\underline x_\perp}(\underline x_{\perp}').
\end{equation}
The following useful identity follows from (\ref{eq:kleingordonwedge}) and
(\ref{eq:conjfield}):
\begin{equation}\label{eq:useful}
\I\left[ L_{\mathrm F}, P(0,u, \underline x_\perp)\right] = -\left(-u\partial_u
u\partial_u + u^2(-\Delta_{\perp} + m^2)\right)Q(0,u,\underline x_\perp).
\end{equation}

For an algebraic formulation of the dynamics, indispensable in what follows,
we need to identify the observable algebra of the theory. The observable
algebra of the field is  $\Af:=\{W(f) | f\in {\cal S}(\R^{d+1}, \R)\}''$, with
$W(f)=\e^{-\I Q[f]}$ the usual Weyl operators. One should think of the
observable algebra as containing all bounded functions of the (smeared) field
operators $Q[f]$ or, more pictorially, all observables that can be constructed
from the $Q(x)$, $x\in \R^{d+1}$. Associated to the right and left wedges are
local algebras of observables ${\mathcal
  A}_{\mathrm F;\mathrm{R, L}}:=\{W(f) | f\in
{\cal S}(W_{\mathrm{R,L}},\R)\}''$. Again, those should be thought of as
containing all observables that can be constructed with the field operators
$Q(x)$, for $x$ belonging to the wedge considered.  As pointed out above, one
can define on ${\mathcal A}_{\mathrm F}$ an automorphism group $\alpha_\tau^0$ by
\begin{equation}\label{eq:freedynamfield}
\alpha_{{\mathrm F},\tau}^0(A)=\e^{\I L_{\mathrm F}\tau} A\ \e^{-\I L_{\mathrm
F}\tau},\qquad A\in{\mathcal A_{\mathrm F}}.
\end{equation}
We note that $\alpha^0_{{\mathrm F},\tau}$ leaves ${\mathcal A}_{\mathrm F;\mathrm{R}}$ invariant.


\subsection{The free detector}\label{s:freedetector}
As pointed out in the introduction, we think of the detector as a two-level
system. Our results extend without problem to an $N$-level system, at the cost
of irrelevant notational complications. So we follow the physics literature on
the subject and limit ourselves to a highly idealized two-level detector. Its
observable algebra is simply the algebra of two by two matrices ${\mathcal
B}(\C^2)$. It will be convenient to use a representation of this algebra in
which both the ground state and the Gibbs state at inverse temperature $\beta$
are represented by vectors. This representation, well known in the
mathematical physics literature on quantum statistical mechanics, is of course
different from the usual one in the standard physics literature in which the
latter is represented by a density matrix.

It is defined as follows. One represents the observable algebra ${\mathcal
B}(\C^2)$  as $\Ad:={\mathcal
  B}(\C^2)\otimes\bbbone_2$ on $\H_{\mathrm
  D}=\C^2\otimes\C^2$, with in particular $A^\dagger:=\left[
\begin{array}{cc} 0&1\\0&0\end{array}\right]\otimes \bbbone_2$. The algebra $\Ad$ is generated by the identity
operator, $A^\dagger$, $A$ and $A^\dagger A$ and one has $A A^\dagger +
A^\dagger A=\bbbone$.

In this representation, the free Heisenberg evolution of the detector with
respect to its proper time $\sigma$ is generated by the self-adjoint operator
$L_{\mathrm D}:=H_{\mathrm D}\otimes \bbbone_2-\bbbone_2\otimes H_{\mathrm
D}$, with
\begin{equation}
H_{\mathrm D}=
\left[\begin{array}{cc} 
E & 0\\ 0 & 0\end{array}\right],
\label{hd}
\end{equation}
for some $E>0$, where $E$ represents the excitation energy of the detector; $L_{\mathrm D}$ is referred to as the
free Liouvillean of the detector. To see this it is enough to remark that
$$
A^\dagger(\sigma):= \alpha_{{\mathrm D},\sigma}^0(A):=\e^{\I L_{\mathrm
D}\sigma} A^\dagger \e^{-\I
  L_{\mathrm D}\sigma}
$$
satisfies  the correct Heisenberg equation of motion
\begin{equation}\label{eq:heisenbergdetector}
\dot A^\dagger(\sigma)=i E A^\dagger(\sigma)
\end{equation}
of an unperturbed two-level system.  Note that the energy levels of the
detector are thought of in this model as pertaining to internal degrees of
freedom (\cite{u,uw, ta}). One should think of the two-level system as being
dragged through spacetime by an external agent that ensures it has constant
acceleration $a$. So the translational degrees of freedom of the detector are
not dynamical variables in this kind of model. In the representation above,
the ground state of the detector can be represented by the vector $|--\rangle$
and the Gibbs state at inverse temperature $\beta$ by the vector
$$
|\beta, \mathrm D\rangle:=(1+\e^{-\beta E})^{-1/2}(|--\rangle + \e^{-\beta
E/2}|++\rangle)\in \Hd.
$$
Indeed, one easily checks that, for any $B\in{\mathcal B}(\C^2)$,
$$
\langle \beta, \mathrm D| B\otimes \bbbone_2|\beta, \mathrm D\rangle
=\frac1{Z_{\beta, \mathrm D}} \mathrm{Tr}\ \e^{-\beta H_{\mathrm D}}B,\quad
Z_{\beta, \mathrm D}=\mathrm{Tr}\ \e^{-\beta H_{\mathrm D}}.
$$
It is the fact that both the ground state and positive temperature states of
the detector can be represented by vectors that makes this representation
particularly suitable for the problem at hand.

\subsection{The uncoupled field-detector system}\label{s:uncoupldsystem}
It is now easy to describe the observable algebra of the joint detector-field
system, as well as its uncoupled dynamics. On the Hilbert space
$\H:=\Hd\otimes \eff$ we consider the observable algebra ${\mathcal
  A}:=\Ad\otimes\Ar$ and the self-adjoint operator $L_0=L_{\mathrm D}\otimes \bbbone_{\mathrm \eff}+
\bbbone_{\Hd}\otimes a L_{\mathrm F}$. The latter determines an automorphism
group
$$
\alpha^0_{\sigma}=\alpha^0_{{\mathrm D}, \sigma}\otimes \alpha^0_{{\mathrm
F},a\sigma}
$$
of $\mathcal A$ in the usual way: $\alpha^0_{\sigma}(B)=\e^{\I L_0\sigma}
B\e^{-\I L_0\sigma}$, $B\in\mathcal A$. Setting
$B(\sigma):=\alpha_\sigma^0(B)$ this yields a solution of the Heisenberg
equations of motion of the uncoupled detector-field system on the Rindler
wedge $W_{\mathrm R}$, which are given by (\ref{eq:kleingordonwedge}) and
(\ref{eq:heisenbergdetector}), with $\tau=a\sigma$.

We will be mostly interested in the state of the system where, initially, the
detector is in its ground state, and the field in its Minkowski vacuum. This
state is represented by the vector $|\mathrm g\rangle:=|--\rangle\otimes
|0\rangle\in \H$. We will write, for any $B\in\mathcal A$:
\begin{equation}\label{eq:groundstate}
\langle B\rangle_{\mathrm g} := \langle \mathrm g| B|\mathrm g\rangle
\end{equation}


\subsection{The coupled field-detector system}
For the coupled system we will use the same representation of the observable
algebra, but change the dynamics. We will give a precise and mathematically
rigorous definition of the dynamics below but to link it with the physics
literature on the subject, we start with a formal computation. Let
$C(\sigma)=[A(\sigma), A^\dagger(\sigma)]$. According to
\cite{uw}, the Heisenberg equations of motion of the observables of the
coupled system are
\begin{eqnarray}\label{eq:heisenbergcoupled}
(\Box + m^2)Q(x)&=&-\lambda\rho( x_* )(A+A^\dagger)(\frac{\tau(x)}{a})\\
\dot A(\sigma)&=&-\I E A(\sigma) +\I\lambda C(\sigma)\int\d u\d \underline
x_\perp a u\ \rho(x_* ) Q(a\sigma, u, \underline x_\perp),\nonumber
\end{eqnarray}
The function $\rho$ tunes the coupling between the detector and the field. It
is evaluated at 
$$
x_*:=x-x(\tau(x)/a),
$$
the spacelike vector linking
$x$ in the right wedge to the instantaneous position $x(\sigma)$ of the
detector, at proper time $\sigma=\tau(x)/a$, where
$\tau(x)$ is the Rindler time coordinate defined in (\ref{eq:coordchange}).

Let $(\tau,u,\underline x_\perp)$ be the Rindler coordinates of the point $x$
then the ones of $x(\sigma)$ are $(\tau,1/a,\underline 0_\perp)$ and hence we
may identify $x_*$, whose coordinates are $(0,u-1/a,\underline x_\perp)$, with an element of 
$(-1/a, +\infty)\times \R^{d-1}$. We take the coupling function $\rho$ to be
in $C_0^\infty( (-1/a,
+\infty)\times \R^{d-1} )$, normalized as $\int \rho(\underline x) \d
\underline x^d=1$.  Typically we imagine $\rho$ to be peaked at the origin, so that the field is
coupled strongest at the position of the detector. Only for such couplings
does it 
make sense to interpret $\sigma$ as the proper time of the detector. Indeed,
if the detector is coupled to the field over a large spatial region,
different parts of the detector undergo a different acceleration and have
a different proper time. The mathematical result we obtain then still
holds, but does no longer have the same physical interpretation. A coupling
{\it strictly} localized at the position of the detector is 
formally given by $\rho(\underline x)=\delta(\underline x)$, a situation
which does not fit the rigorous mathematical setup presented in this
work. We will comment further on the role played by the choice of coupling in
Section \ref{sectfgrc}.

Using  (\ref{eq:equaltime}) and (\ref{eq:useful}), it is easy to
show  through a formal computation that these equations are satisfied by the
operators $Q^{(\lambda)}(\tau, u, \underline x_\perp)$ and
$A^{(\lambda)}(\sigma)$ defined as follows:
$$
Q^{(\lambda)}(\tau, u, \underline x_\perp):=\e^{\i \tilde
L_\lambda\frac{\tau}{a}}\ Q(0,u, \underline x_\perp)\ \e^{-\i \tilde
L_\lambda\frac{\tau}{a}},\quad  A^{(\lambda)}(\sigma):=\e^{\i \tilde
L_\lambda\sigma}\ A\ \e^{-\i \tilde L_\lambda\sigma},
$$
where
\begin{equation}\label{eq:coupledliouvillean}
\tilde L_\lambda:=L_0 +\lambda I, I:=(A+A^\dagger) \int \d u\d \underline
x_\perp au\ \rho(x_*|_{\tau=0})Q(0,u,\underline x_\perp)
\end{equation}
and $x_*|_{\tau=0}$ is given in Rindler coordinates by $(0,u-1/a,\underline x_\perp)$. In
other words, the Liouvillean $\tilde L_\lambda$ generates the correct
Heisenberg dynamics of the observables in the representation at hand. 

{\it Remark. } The analysis we carry out in this paper works for general interactions of the form $I =
G\cdot Q(g) + G^*\cdot Q(\overline g)$, and for sums of such terms, where $G$
are matrices acting on the detector space, and $g\in L^2({\mathbb R}^3,
\d\underline x)$ are ``form factors''. 

The following result is proved in Section \ref{dynpropproofsect}.

\begin{proposition}
\label{dynproposition}
The operator $\tilde L_\lambda$ in (\ref{eq:coupledliouvillean}) is for all
$\lambda$  essentially
self-adjoint on $D(L_0)\cap D(I)$ and the maps $\alpha^\lambda_\sigma(B):=  \e^{\I \tilde L_\lambda
\sigma}B \e^{-\I \tilde L_\lambda\sigma}$ with $\sigma\in\R$ and $B\in\mathcal
A$ define a weakly continuous one-parameter group of automorphisms of the
observable algebra $\mathcal A$.
\end{proposition}


\subsection{The result}\label{s:result}

We are now in a position to give a precise statement of our result. Define
\begin{equation} 
g(\varkappa,\underline k_\perp) 
=
\widehat{\Big(x_1\rho(x_*|_{\tau=0})\Big)}\big((|\underline k_\perp|^2+m^2)^{1/2} \sinh\varkappa,\underline k_\perp\big),
\label{m7}
\end{equation}
where $\widehat{\ }$ denotes the Fourier transform.
\begin{theorem} 
\label{thm:partialresult} 
Let $d\geq 1$ if $m>0$ and $d\geq 2$ if $m=0$, and suppose the following ``Fermi Golden Rule Condition'' holds, 
\begin{equation}
\label{fgrc}
\int_{{\mathbb R}} \d \varkappa \ \e^{-\I \frac Ea \varkappa}\
g(\varkappa,\underline k_\perp)\neq 0 
\mbox{\ \ \ \  for some $\underline k_\perp\in {\mathbb R}^{d-1}$}.
\end{equation}
Then there is a constant $\lambda_0>0$ s.t. if $0<|\lambda|<\lambda_0$ then 
\begin{equation}
\lim_{\sigma\to\infty}\langle\alpha_\sigma^\lambda(B)\rangle_{\mathrm g}
=\frac{1}{Z_{\beta,\mathrm D}} \mathrm{Tr}\ \e^{-\beta
  H_{\mathrm D}}B+ O(\lambda^2),
\label{mm1}
\end{equation}
for all  $B\in \mathcal B(\C^2)$, and where $\beta= \frac{2\pi}{a}$.

More generally, if $\varrho$ is any density matrix on $\H$ then
\begin{equation}
\lim_{\sigma\to\infty}{\mathrm Tr}\, \varrho\alpha_\sigma^\lambda(B F)
=\Big(\frac{1}{Z_{\beta,\mathrm D}} \mathrm{Tr}\ \e^{-\beta
  H_{\mathrm D}}B\Big)\, \langle 0| F|0\rangle+ O(\lambda^2),
\label{mm0}
\end{equation}
for any detector observable $B\in \mathcal B(\C^2)$ and any field observable
$F\in {\cal A}_{\mathrm F}$. 
\end{theorem}
Result (\ref{mm1}) shows that if at $\sigma=0$ the detector-field system is in a state which
is a local perturbation of its ground state, then the reduced density matrix
of the detector converges asymptotically in time to the detector's Gibbs state
at inverse temperature $\beta= \frac{2\pi}{a}$. This is a (slightly) stronger
statement than the formulations usually found in the literature, since it
allows both the field and the detector to be initially in an excited state.

{\it Remarks.\ }
1) Theorem \ref{thm:partialresult} follows from a more complete result, stated
as Theorem \ref{thm:fullresult} below, where the r.h.s. of (\ref{mm1}) is identified as the equilibrium state
$|\lambda\rangle\in{\cal H}$ of the coupled system, see also (\ref{eq:kmscoupled}) below. An expansion of
$\langle\lambda|\cdot|\lambda\rangle$ for small $\lambda$ yields the uncoupled equilibrium state
plus an error of second order in $\lambda$ (the absence of a first order error
term is due to the fact that the expectation of the interaction $I$ in the
uncoupled equilibrium state vanishes). 

2) The approach to the limit state in (\ref{mm1}) is exponentially fast, 
$$
\left| 
{\mathrm Tr}\, \varrho\alpha_\sigma^\lambda(B)
 - \langle \lambda | B|\lambda\rangle\right| < C\|B\| \e^{-\lambda^2\eta
 \sigma},
$$
where $C$ is a constant (depending on the interaction, but not on the initial
density matrix $\varrho$ nor on $B$) and $\eta = (1+\e^{-2\pi E/a})\xi$, with 
\begin{equation}
\xi \equiv\xi(E) = \frac{1}{2a}\int_{{\mathbb R}^{d-1}}\d\underline k_\perp  \left|\int_{\mathbb
    R}\d\varkappa \ \e^{-\I \frac Ea \varkappa}g(\varkappa,\underline
  k_\perp)\right|^2\geq 0.
\label{m19}
\end{equation}
The quantity $\tau_{\mathrm{relax}}=1/\lambda^2\eta$ is called the {\it
  relaxation time} of the process. 
The purpose of condition (\ref{fgrc}) is to ensure that $\xi>0$, i.e., that
$\tau_{\mathrm{relax}}<\infty$. We will show in the following subsection that
this is typically the case.

We finally remark that, whereas the leading term of the right hand side of (\ref{mm1}) does not
depend on the choice of form factor $\rho$ in the interaction term, the
relaxation time $\tau_{\mathrm{relax}}$ does, via (\ref{m7}) and (\ref{m19}). Nevertheless, we show in
the next subsection that $\tau_{\mathrm{relax}}$ is independent of the form
factor for interactions sharply localized at the position of the detector.

\subsection{The Fermi Golden Rule Condition}
\label{sectfgrc}

The goal of this section is to show that (\ref{fgrc}) is satisfied for
``generic'' interactions. 
\begin{proposition}
\label{fgrcprop}
Take the coupling function $\rho$ in (\ref{eq:heisenbergcoupled}),
(\ref{eq:coupledliouvillean}) to be of the form
$\rho(\underline x)
= \rho_1(x_1)\rho_\perp(\underline x_\perp)$, (``square detector'') with
$\rho_1\geq 0$. Then condition 
(\ref{fgrc}) is satisfied for all $E$ except for $E\in{\cal E}$,
where $\cal E$ is a discrete (possibly empty) subset of $\mathbb R$. In
particular, $\xi(E)>0$ for all $E\not\in\cal E$.
\end{proposition}

Values of $E$ satisfying $\xi(E)=0$ (which form necessarily a subset of $\cal E$ in the proposition) correspond to energy gaps of the detector
Hamiltonian for which thermalization of the detector occurs (if at all) with a
larger relaxation time at least of the order $\lambda^{-4}$ (as opposed to 
$\lambda^{-2}$ for $E$ s.t. $\xi(E)>0$), see \cite{MLSO}.

For a particular choice of the coupling function $\rho$ one may
resort to a numerical study of the condition (\ref{fgrc}). On the analytic
side we can calculate $\xi$, (\ref{m19}), in the limit of a strictly localized
interaction. More precisely, we choose $\rho_1$, $\rho_\perp$ as in
Proposition \ref{fgrcprop}, and consider the family $\rho_\epsilon(\underline x)
= \epsilon^{-d}\rho_1(x_1/\epsilon)\rho_\perp(\underline x_\perp/\epsilon)
\rightarrow \delta(x_1-1/a)\delta(\underline x_\perp)$
which represents an interaction localized exactly at the position of
the detector in the limit $\epsilon\rightarrow 0$. Each $\epsilon$ defines
thus a $\xi_\epsilon(E)$ by (\ref{m19}), and we obtain, for $d=3$ and $m>0$, 
\begin{eqnarray*}
\lefteqn{
\lim_{\epsilon\rightarrow 0}\xi_\epsilon(E)}\\
&=& 
\frac a2 \int_{{\mathbb R}^2}\frac{\d\underline k_\perp}{\omega_\perp^4}\left| \int_{\mathbb R}\d\varkappa
\frac{2\sinh^2\varkappa-\frac{E}{\omega_\perp}\cosh\varkappa-1}{(\frac{E}{\omega_\perp} +\cosh\varkappa)^4} \e^{-\i\big[\frac Ea \varkappa +\frac{\omega_\perp}{a}
\sinh\varkappa\big]}\right|^2,
\end{eqnarray*}
where $\omega_\perp=\sqrt{|\underline k_\perp|^2+m^2}$. This limit does not
depend on the form of $\rho$ and the leading term of $\tau_{\mathrm{relax}}$
(as $\epsilon\rightarrow 0$) is thus independent of the detector form factor.

\medskip

{\it Proof of Proposition \ref{fgrcprop}.\ }
We denote the integral in (\ref{fgrc}) by $\I
\widehat{\rho_\perp}(\underline k_\perp) J(E,\omega_\perp)$,
where $\omega_\perp=\sqrt{|\underline k_\perp|^2+m^2}$, see also
(\ref{m7}). For $\omega_\perp\neq0$ we can make the change of variable
$y=\omega_\perp \sinh\varkappa$ to obtain the representation 
\begin{equation}
J(E,\omega_\perp) = 
\int_{\mathbb R}\d
  y \frac{\e^{-\I \frac Ea \mathrm{argsinh}(y/\omega_\perp)}}{\sqrt{\omega_\perp^2 + y^2}} f(y),\ \ \ f(y):=
\e^{-\I y/a}\Big(\frac{-\I}{a}\widehat{\rho_1}(y) + \widehat{\rho_1}'(y)\Big).
\label{u1}
\end{equation}
We view $\omega_\perp^2=\mu$ in the integral as a parameter, $\mu>0$. We first
show that given any $\mu_0>0$, the integral in (\ref{u1}), for $E=0$, does
not vanish identically in any neighbourhood of $\mu_0$. 

Let us consider $\mu_0=1$; a simple modification of the following argument
yields the general case. Assume {\it ad absurdum} that $J(0,\mu)=0$
for all $\mu$ in a neighbourhood of $1$. Then, by taking derivaties
of $J(0,\mu)$ with respect to $\mu$, at $\mu_0=1$,
we see that 
\begin{equation}
\int_{\mathbb R}\d y \ (1+y^2)^{-n}\, (1 + y^2)^{-1/2} f(y)=0,
\label{u2}
\end{equation}
for all $n=0,1,\ldots$  Now, it is not difficult to verify that the linear span
of all functions $(1+y^2)^{-n}$, $n=1,2,\ldots$ is dense in the space of even
functions in $L^2({\mathbb R},dy)$. (One may prove this with little effort
via the Fourier transform, for example.) It thus follows from (\ref{u2}) that
the even part of $f$ must vanish, 
$f(y)+f(-y)=0$ for all $y\geq 0$. In particular, $f(0)=0$, which means that 
\begin{equation}
a^{-1} =-\I \widehat{\rho_1}'(0)
\label{u3}
\end{equation}
(we assume without loss of generality that $\rho_1$ is normalized as
$\int_{\mathbb R}\d x\, \rho_1(x)=1$). On the other hand, we have $-\I
\widehat{\rho_1}'(0) = -\int_{\mathbb R}\d x \, x\rho_1(x)<a^{-1}$, since in
the integral, $x>-a^{-1}$ due to the fact that $\rho_1$ is supported in
$(-1/a,\infty)$. Therefore condition (\ref{u3}) is not verified. 

This shows that given any $\mu_0>0$ we can find a $\mu_1>0$ (arbitrarily close
to $\mu_0$) with the property that $J(0,\mu_1)\neq 0$. 

Pick a nonzero $K_0\in {\mathbb R}^{d-1}$ satisfying
$\widehat{\rho_\perp}(K_0)\neq 0$ and set $\mu_0:=\sqrt{|K_0|^2+m^2}$. Then,
by the above argument and by the continuity of $\widehat{\rho_\perp}$ there is 
a $\mu_1=:\sqrt{|K_1|^2+m^2}$ (which is close to $\mu_0$ and defines a $K_1$
close to $K_0$) s.t. $J(0,\mu_1)\neq 0$ {\it and}
$\widehat{\rho_\perp}(K_1)\neq 0$. Hence we have shown that there exists a
nonzero $K_1$ satisfying $\I \widehat{\rho_\perp}(K_1) J(0,\omega_1)\neq 0$,
where $\omega_1=\sqrt{|K_1|^2+m^2}$. Condition (\ref{fgrc}) is thus satisfied
for $E=0$.

Finally we pass to the other values of $E$ by an analyticity argument. Indeed,
one easily sees (best by using the form of $J$ in which one integrates over
$\varkappa$ rather than $y$, c.f. (\ref{m7}), (\ref{fgrc})) that the map $E\mapsto J(E,\omega_1)$ is analytic and by
the previous argument it does not vanish at $E=0$. Thus the zeroes of this
map are contained in a discrete set ${\cal E}(\omega_1)\subset{\mathbb
  C}$. Any $E$ avoiding this set thus satisfies (\ref{fgrc}).
\hfill $\blacksquare$

\section{Proof of Theorem \ref{thm:partialresult}}\label{s:proof}
\subsection{Strategy}
As mentioned in the introduction, the first ingredient of the proof is the
observation that the Minkowski vacuum is (a realization of) the GNS
representation of a KMS state on the right wedge algebra for the Lorentz
boosts at the inverse temperature $\beta=2\pi$. This is the content of
Theorem \ref{thm:kmsfield} below. To give a precise statement, we need the
so-called modular conjugation operator, defined as follows:
\begin{equation}
J_{\mathrm F}=\Gamma(j_{_{\mathrm F}}),\ {\mathrm {where}}\ \forall \ \psi \in L^2(\R^d, \d
\underline x), \ j_{_{\mathrm F}}\psi(\underline x)=\overline \psi(-x_1, \underline x_\perp);
\label{m10}
\end{equation}
here $\Gamma(j)$ stands for the second quantization of $j$.

\begin{theorem}[\cite{biwi}] 
\label{thm:kmsfield} The Fock vacuum in
$\eff$ induces on $\mathcal A_{\mathrm{F,R}}$ a state which is KMS at inverse
temperature $\beta=2\pi$ for $\alpha_{\mathrm F,\tau}^0$. In particular, one
has
$$
\Ar'=\Al,\quad J_{\mathrm F}\Ar J_{\mathrm F}=\Al,\quad
\overline{\Ar|0,\mathrm{F}\rangle}=\eff=\overline{\Al|0,\mathrm F\rangle},
$$
and for all $f\in {\cal S}(W_{\mathrm R},{\mathbb C})$: ${ L_{\mathrm F}|0,
  \mathrm F\rangle=0}$ and  $\e^{-\pi L_{\mathrm
F}}Q[f]|0,\mathrm F\rangle=J_{\mathrm F}Q[\overline f]|0,\mathrm F\rangle
$.  
\end{theorem}
This result was proven in considerable generality in \cite{biwi}, for
relativistic fields satisfying the Wightman axioms. The result above for the free
scalar field can be obtained from essentially direct computations, and we shall not detail
it.

Similarly, the states $|\beta, \mathrm D\rangle$ introduced in Section
\ref{s:freedetector} are GNS representatives of the KMS  states at inverse
temperature $\beta$ for the free detector dynamics $\alpha_{\mathrm D,
\sigma}$ on the detector observable algebra ${\mathcal B}(\C^2)$. This well
known observation is for convenience summarized in the following lemma.  The
appropriate conjugate operator is given by
$$
J_{\mathrm D}=E(C\otimes C),
$$
where  $C$ is the antilinear operator of complex conjugation on $\C^2$ and $E$
is the exchange operator on $\C^2\otimes\C^2$, $E\varphi\otimes\chi=\chi\otimes\varphi$.
\begin{lemma} For any $\beta>0$, the vector $|\beta, \mathrm D\rangle$ induces on ${\mathcal B}(\C^2)$ a state that is KMS at inverse temperature $\beta$ for $\alpha_{\mathrm D, \sigma}^0$. In particular, one has
$$
\Ad'=\bbbone_2\otimes \mathcal B(\C^2),\quad \Ad'=J_{\mathrm D}\Ad J_{\mathrm
  D},\quad
\Ad|\beta, D\rangle =\H=\Ad'|\beta, D\rangle,
$$
and
$$
\e^{-\beta L_{\mathrm D}/2}(B\otimes \bbbone_2)|\beta, \mathrm D\rangle
=J_{\mathrm D} (B^*\otimes \bbbone_2)|\beta, \mathrm D\rangle.
$$
\end{lemma}
Defining, on $\H=\Hd\otimes\eff$, $J:=J_{\mathrm D}\otimes J_{\mathrm F}$,
it follows that the vector
\begin{equation}
\label{eq:kmsuncoupled}
|\mathrm 0\rangle:= |\beta =2\pi/a\rangle \otimes |0,\mathrm F\rangle
\end{equation}
is a GNS representative of the KMS state at inverse temperature $\beta=\frac{2\pi}{a}$
for the free dynamics $\alpha_\sigma^0$ on $\A=\Ad\otimes\Ar$. This  suggests 
to treat the problem at hand as one of return to
equilibrium.

The rest of the argument then proceeds in three steps:

(a) One proves the existence of a GNS representative $|\lambda\rangle\in\H$,
defined below,
of the KMS state for the perturbed dynamics at the
same temperature (Section \ref{s:perturbation});

(b)  One reduces the proof of Theorem \ref{thm:fullresult} and hence of
Theorem \ref{thm:partialresult} to showing that
the generator of the perturbed dynamics has a
simple eigenvalue at $0$ and otherwise absolutely continuous spectrum only;

(c) One finally uses  spectral deformation theory to prove these two statements.

The strategy in (a)-(b)-(c) has been applied successfully to radiative problems
in atomic physics, the spin-boson model, and similar systems in \cite{JP1,JP2,BFS,M,DJP}, where we refer for further
references. A concise introduction to the field can be found in \cite{pi}. The
implementation of this strategy in the present context is reasonably
straightforward. We will detail those points that are specific to the
current situation.


\subsection{Perturbation theory}\label{s:perturbation}
We define on $\H=\H_{\mathrm D}\otimes \eff$,
in addition to $\tilde L_\lambda$ (see \ref{eq:coupledliouvillean}), the
so-called {\it standard Liouvillean}
\begin{equation}\label{eq:standardliouvillean}
L_\lambda =\tilde L_\lambda -\lambda JIJ. 
\end{equation}
We outline the proof of the following result in Section
\ref{dynpropproofsect}. 
\begin{lemma} 
\label{lem:standardliouvillean} $L_\lambda$ is essentially
  self-adjoint on $D(L_0)\cap D(I)\cap D(JIJ)$  and, for all $B\in\A$,
$$
\alpha_\sigma^\lambda(B) =\e^{\i L_\lambda \sigma}B\e^{-\i L_\lambda \sigma}.
$$
\end{lemma}

A useful feature of the standard Liouvillean (in fact, the motivation for its
definition!) is that the
unitary it generates leaves the equilibrium state of the coupled system
invariant, see (\ref{eq:kernel}) below.
\begin{proposition} 
\label{prop:perturbedkms} 
The vector $|0\rangle$ representing the uncoupled equilibrium state,
(\ref{eq:kmsuncoupled}), is in the domain of the unbounded operator
$\e^{-\frac{\pi}{a}\tilde L_\lambda}$, and the vector 
\begin{equation}
\label{eq:kmscoupled}
|\lambda\rangle := \frac{\e^{-\frac{\pi}{a}\tilde L_\lambda}|0
\rangle}{\|\e^{-\frac{\pi}{a}\tilde L_\lambda}|0\rangle\|}\in\H
\end{equation}
defines a $(\frac{2\pi}{a}, \alpha_\sigma^\lambda)$-KMS state on $\A=
\Ad\otimes \Ar$ and it satisfies
\begin{equation}\label{eq:kernel}
L_\lambda |\lambda\rangle =0.
\end{equation}
\end{proposition}

{\it Proof. } To show that $|0\rangle \in\dom(\e^{-\frac{\pi}{a}\tilde
  L_\lambda})$ we check that the Dyson series
\begin{equation}
\sum_{n\geq 0}(-\lambda)^n \int_0^{\pi/a}\d t_1\cdots \int_0^{t_{n-1}}\d t_n\ 
\alpha_{\I t_n}^0(I)\cdots \alpha_{\I t_1}^0(I) |0\rangle
\label{m1}
\end{equation}
converges. We write the interaction operator conveniently as $I = G Q[g]$, 
where $G=A+A^\dagger$ and $g(x)=a\delta(x^0)x^1\rho(x_*(\underline
x))$ has support in $W_{\mathrm R}$,
c.f. (\ref{eq:coupledliouvillean}). In (\ref{m1}) we have set, for real $s$,
$$
\alpha_{\I s}^0(I) = \e^{-s L_{\mathrm D}}G\e^{s L_{\mathrm D}} \ 
\e^{-s L_{\mathrm F}} Q[g] \e^{s L_{\mathrm F}}.
$$
To see that $\e^{-s L_{\mathrm F}} Q[g] \e^{s L_{\mathrm F}}$ is well defined
for $0\leq s\leq\pi/a$ one shows that since $g$ is supported in the right
wedge, the map $t\mapsto \e^{\I t L_{\mathrm F}} Q[g] \e^{-\I t L_{\mathrm F}}
= Q[g\circ B_{-t}]$ has an analytic continuation into the strip $0<{\rm Im}\,
t<\pi/a$, and it is continuous at the boundary of the strip ($t\in{\mathbb
  R}$, $t\in \I\frac{\pi}{a}{\mathbb R}$). This argument is actually part of
the proof of the Bisognano--Wichmann theorem, \cite{biwi}. It follows in
particular that the integrals in (\ref{m1}) are well defined and that
furthermore 
$$
\sup_{0\leq {\rm Im}\, s \leq \pi/a} \left\| \alpha_{\I s}^0(I)
  (N+1)^{-1/2}\right \| = C<\infty,
$$
where $N$ is the number operator on Fock space. Since $|0\rangle$ is the
vacuum on the field part, and each interaction term
$\alpha_{\I s}^0(I)$ can increase the particle number by at most one we have
the bound $\|\alpha_{\I t_n}^0(I)\cdots \alpha_{\I t_1}^0(I) |0\rangle\|\leq
C^n\sqrt{n!}$. It follows that the series (\ref{m1}) converges (for all values
of $\lambda$) and hence $|0\rangle \in\dom(\e^{-\frac{\pi}{a}\tilde
  L_\lambda})$. 

The facts that $|\lambda\rangle$ defines a
$(\frac{2\pi}{a},\alpha_\sigma^\lambda)$-KMS state and that $L_\lambda
|0\rangle =0$ follow from Araki's perturbation theory of KMS states, and from
perturbation theory of standard Liouville operators, see \cite{DJP}. \hfill
$\blacksquare$ 

We are now in a position to state the full result, of which Theorem \ref{thm:partialresult}
is an immediate consequence:
\begin{theorem}
\label{thm:fullresult} 
Assume that the Fermi Golden Rule Condition (\ref{fgrc}) is satisfied. There exists $\lambda_0$ so that for all
$0<|\lambda|< \lambda_0$, for all density matrices $\varrho$ on $\H$ and for
all $B\in\A$
$$
\lim_{\sigma\to\infty} {\mathrm{Tr}}\ \varrho \ \alpha_\sigma^\lambda(B) =
\langle \lambda|B|\lambda\rangle.
$$
\end{theorem}

{\it Proof.} We show in Section \ref{redsection} that the result follows if
the spectrum of $L_\lambda$ is purely absolutely continuous
with the exception of a single simple eigenvalue at zero. These spectral
characteristics are shown in Theorem \ref{specthm}. \hfill $\blacksquare$


\subsection{Reduction to a spectral problem}
\label{redsection}
We reduce the proof of Theorem \ref{thm:fullresult} to a spectral problem
via the following simple lemma, which is a variant of the
Riemann-Lebesgue lemma:
\begin{lemma} 
\label{stephansfamouslemma}Let $\H$ be a Hilbert space, $\phi\in\H$, $\mathcal A$
a subalgebra of $\mathcal B(\H)$ whose commutant we denote by ${\cal A}'$, and
let $L$ be a self-adjoint operator on $\H$. Suppose that ${\mathcal A'\phi}$
is dense in $\H$, that $\e^{\I L \tau} \mathcal A \e^{-\I L \tau}\subset
\mathcal A,\ \forall \tau$, that $L\phi=0$, and that on the orthogonal
complement of $\phi$, $L$ has purely absolutely continuous spectrum.

Then we have 
\begin{equation}
\lim_{\tau \to \infty}\mathrm{Tr}\ \varrho\ \e^{\I L \tau} B \e^{-\I L
\tau}=\langle \phi, B\phi\rangle,
\label{m2}
\end{equation}
for all $A\in{\cal A}$ and for all density matrices $0\leq \varrho \in \mathcal
L^1(\H)$, ${\mathrm Tr}\varrho=1$.
\end{lemma}

{\it Proof.} We may diagonalize $\varrho=\sum_{n=1}^\infty p_n
|\psi_n\rangle\langle\psi_n|$, where $\psi_n\in{\cal H}$ and the probabilities
$0\leq p_n\leq 1$ sum up to one. So it suffices to show (\ref{m2}) for a
rank-one density matrix $\varrho=|\psi\rangle\langle\psi|$. Given any
$\epsilon>0$ there is a $B'\in {\cal A}'$
s.t. $\|\psi-B'\phi\|<\epsilon$. Thus by replacing $\psi$ by $B'\phi$, 
commuting $B'$ and $\e^{\I L \tau} A \e^{-\I L \tau}$, and by using the
invariance of $\phi$ under $\e^{-\I L \tau}$ we obtain
\begin{equation}
{\mathrm Tr\,}\varrho\,\e^{\I L \tau} B \e^{-\I L \tau}= \langle\psi,\e^{\I L \tau} B \e^{-\I L \tau}\psi\rangle=\langle \psi, B'
\e^{\I L \tau}B\phi\rangle +O(\epsilon),
\label{m3}
\end{equation}
where the remainder is estimated uniformly in $\tau$. Since the spectrum of
$L$ is absolutely continuous except for a simple eigenvalue at zero with
eigenvector $\phi$, the propagator $\e^{\I L\tau}$ converges in the weak sense
to the rank-one projection $|\phi\rangle\langle\phi|$, as
$\tau\rightarrow\infty$. Using this in (\ref{m3}), together with the facts that
$\langle\psi,B'\phi\rangle = 1+O(\epsilon)$, and that $\epsilon$ can be chosen
 arbitrarily small yields relation (\ref{m2}). \hfill $\blacksquare$

We apply Lemma \ref{stephansfamouslemma} with $L=L_\lambda$ and 
$\phi=|\lambda\rangle$. The density of ${\cal A}'|\lambda\rangle$ follows
from the KMS property of $|\lambda\rangle$, the invariance of ${\cal A}$ under
 $\e^{\I L_\lambda\tau}\cdot \e^{-\I L_\lambda\tau}$ follows from Lemma
\ref{lem:standardliouvillean} and the relation $L_\lambda|\lambda\rangle=0$ is
shown in Proposition \ref{prop:perturbedkms}. It remains to prove that on the orthogonal complement of $|\lambda\rangle$, $L_\lambda$ has purely
absolutely continuous spectrum.


\section{Spectral analysis of $L_\lambda$}
\label{specanal}

The spectrum of the operator $L_{\mathrm D}$ consists of two simple eigenvalues
$\pm E$ (eigenvectors $|\pm,\mp\rangle$) and a doubly degenerate eigenvalue at
$0$ (eigenvectors $|\pm,\pm\rangle $). $L_{\mathrm F}$ has absolutely
continuous spectrum covering the entire real axis, and a single embedded
eigenvalue at the origin. This eigenvalue is simple and has eigenvector
$|0,\mathrm F\rangle$. It follows that $L_0$ has absolutely continuous
spectrum covering the axis and three embedded eigenvalues at $0,\pm E$, the
one at $0$ being doubly degenerate.

Our goal is to show that the nonzero eigenvalues are unstable under the
perturbation $\lambda(I-JIJ)$, and that the degeneracy of the eigenvalue zero
is lifted. We do this via {\it spectral deformation theory}, showing that the
unstable (parts of the) eigenvalues turn into {\it resonances} located in the
lower complex plane. 

\subsection{Spectral deformation}
For the spectral analysis it is useful to consider the unitarily transformed
Hilbert space $L^2({\mathbb R}^d,\d\varkappa\,\d^{d-1}\underline k_\perp)$ of
one-particle wave functions of the field, determined by $L^2({\mathbb
  R^d,\d^d\underline x})\ni f\mapsto Wf$ with
\begin{equation}
(Wf)(\varkappa,\underline k_\perp):= \sqrt{\omega_\perp\cosh\varkappa}\ 
\widehat{f}(\omega_\perp\sinh\varkappa,\underline k_\perp),
\label{m4}
\end{equation}
where $\omega_\perp := \sqrt{|\underline k_\perp|^2 +m^2}$ and where
$\widehat{f}$ is the Fourier transform of $f$. The advantage of this
representation of the Hilbert space is that the operator $K$, defined in
(\ref{eq:liouvfreefield}), takes the particularly simple form $K = \I\partial_\varkappa$.
The transformation $W$ lifts to Fock space in the usual way. We do not
introduce new names for spaces and operators in the transformed system. The
 Liouville operator (\ref{eq:coupledliouvillean}) is 
\begin{eqnarray}
L_\lambda &=& L_{\mathrm D}+ L_{\mathrm F}+\lambda V,\nonumber\\
L_0 &=& L_{\mathrm D}+ a L_{\mathrm F},\ \ L_{\mathrm D}= H_{\mathrm
  D}\otimes\bbbone_2 - \bbbone_2\otimes H_{\mathrm D}, \ \ L_{\mathrm F}=\d\Gamma(\I\partial_\varkappa),
\label{m5}\\
V &=& I- JIJ,\ \ \ I=G \otimes\bbbone_2\otimes\frac{a}{\sqrt 2}\left\{ a^\dagger(g) +
  a(g)\right \}
\label{m6}
\end{eqnarray}
acting on the Hilbert space ${\cal H}= {\mathbb C}^2\otimes {\mathbb
  C}^2\otimes{\cal F}$, where ${\cal F}$ is the bosonic Fock space over
$L^2({\mathbb R}^d,\d\varkappa\,\d^{d-1}\underline k_\perp)$. In (\ref{m6})
$G$ is the $2 \times 2$ matrix with $0$ on the diagonal and $1$ on the
off-diagonals, and $g(\varkappa,\underline k_\perp) 
= \big(W \Omega^{-1/2}x_1\rho(x_*|_{\tau=0})\big)(\varkappa,\underline
k_\perp)$ is given in (\ref{m7}).  
The action of $j_{_{\mathrm F}}$, (\ref{m10}), is given by $\big(j_{_{\mathrm
    F}}f\big)(\varkappa,\underline k_\perp)= \overline{f}(-\varkappa,\underline k_\perp)$. 

We describe now the complex deformation. Let $\theta\in {\mathbb R}$. The map
$$
\psi_\theta(\varkappa_1,\ldots,\varkappa_n):=\big(U_\theta\psi\big)(\varkappa_1,\ldots,\varkappa_n)
:=
\e^{\I\theta(\varkappa_1+\cdots \varkappa_n)}\psi(\varkappa_1,\ldots,\varkappa_n)
$$
defines a unitary group on $\cal F$ (we are not displaying the variables
$\underline k_\perp$ in the argument of $\psi$ since $U_\theta$ does not act on
them). An easy calculation shows that
\begin{equation}
L_\lambda(\theta):=U_\theta L_\lambda U_\theta^* = L_0(\theta) +\lambda V(\theta),\ \ 
L_0(\theta) = L_0 -a \theta N, \ \ V(\theta)= I(\theta) - J I(\theta) J,
\label{m8}
\end{equation}
where $N=\d\Gamma(\bbbone)$ is the number operator on $\cal F$ and 
\begin{equation}
I(\theta) = G \otimes\bbbone_2\otimes\frac{a}{\sqrt 2}\left\{ a^\dagger(\e^{\I\theta\varkappa}g) +
  a(\e^{\I\overline\theta\varkappa}g)\right \},
\label{m11}
\end{equation}
where we have put the complex conjugate $\overline \theta$ in the argument of
the annihilation operator in (\ref{m11}) in view of the complexification of
$\theta$. 
\begin{lemma}
\label{analyticityinteraction}
Let
\begin{equation}
\theta_0(m,d):=
\left\{
\begin{array}{cl}
\infty & \mbox{if $m\neq 0$ and $ d\geq 1$}\\
\frac{d-1}{2}  &  \mbox{if $m=0$ and $ d\geq 2$}
\end{array}
\right.
\label{m12}
\end{equation}
where  $m\geq 0$ is the mass of the field and $d$ is the spatial
dimension. We have  $\e^{
\I\theta\varkappa} W\Omega^{-1/2}h \in L^2({\mathbb R}^d,d\varkappa d\underline
k_\perp)$ for all $\theta\in {\mathbb C}$ satisfying $|\theta|<\theta_0$  and for all
$h\in{\cal S}({\mathbb R}^d,{\mathbb C})$. Moreover, for $|\theta|<\theta_0$, $L_\lambda(\theta)$ is a closed operator on the dense domain ${\cal D}=\dom(L_0)\cap\dom(N)$.
\end{lemma}

{\it Proof.} $L_0(\theta)$ is a normal operator, so it is
closed. Assume we know that $\e^{\I\theta\varkappa} W \Omega^{-1/2}{\cal S}({\mathbb R}^d,{\mathbb C})\subset L^2({\mathbb R}^d,d\varkappa d\underline
k_\perp)$, and recall that $x_1\rho(x_*|_{\tau=0})\in {\cal S}({\mathbb R}^d,{\mathbb C})$. Then, for ${\rm Im}\theta\neq 0$ the perturbation $V(\theta)$ is
infinitesimally small w.r.t. $L_0(\theta)$, so $L_\lambda(\theta)$ is closed
by stability of closedness. For ${\rm Im}\theta= 0$ the operator
$L_\lambda(\theta)$ is even selfadjoint. 

Let $h\in{\cal S}({\mathbb R}^d,{\mathbb C})$. According to (\ref{m4}) we have 
$$
(W \Omega^{-1/2}h)(\varkappa,\underline k_\perp) =
\widehat{h}(\omega_\perp\sinh\varkappa,\underline k_\perp).
$$
Since $\widehat{h}\in \cal S$ we have that for any
integer $n$ there is a constant $C_n$ s.t.
$$
\left|\widehat{h}(\omega_\perp\sinh\varkappa,\underline k_\perp)\right| <
\frac{C_n}{1+[m^2\sinh^2\varkappa +|\underline
  k_\perp|^2\cosh^2\varkappa]^n}.
$$
For $m=0$ we thus obtain (using an obvious change of variables) the estimate
\begin{equation}
\int_{\mathbb R} \d\varkappa\ \e^{2\theta'|\varkappa|}\int_{{\mathbb
    R}^{d-1}}\d\underline k_\perp \
\left|\widehat{h}(\omega_\perp\sinh\varkappa,\underline k_\perp)\right|^2 <
\widetilde{C}_n\int_{\mathbb R} \d\varkappa\ \frac{\e^{2\theta'|\varkappa|}}{[\cosh\varkappa]^{d-1}}
\label{m13}
\end{equation}
which is finite provided $\theta'=|{\mathrm Im}\theta|<(d-1)/2$. If $m\neq 0$
then the l.h.s. of (\ref{m13}) is bounded from above by
$$
\int_{\mathbb R} \d\varkappa\int_{{\mathbb R}^{d-1}}\d\underline k_\perp
\ \e^{2\theta'|\varkappa|} \frac{C_n^2}{[1/2+m^2\sinh^2\varkappa]^n [1/2+|\underline
  k_\perp|^2]^n}
$$
which is finite if $\theta'=|{\mathrm Im}\theta|<n$, and $n$ can be
chosen arbitrarily large. \hfill $\blacksquare$

\subsection{Spectra of $L_\lambda(\theta)$ and of $L_\lambda$}

The goal of this section is to prove the following result.
\begin{theorem}
\label{specthm}
Suppose the Fermi Golden Rule Condition (\ref{fgrc}) holds. There is a
$\lambda_0>0$ s.t. if $0<|\lambda|<\lambda_0$ then the spectrum of $L_\lambda$
consists of a simple eigenvalue at zero and is purely absolutely continuous on
the real axis otherwise.
\end{theorem}

Remark that the spectrum of $L_0(\theta)=L_0 -a \theta N$ consists of the {\it isolated
  eigenvalues} $\pm E$ (simple) and $0$ (doubly degenerate), and of the lines
of continuous spectrum $\{{\mathbb R}-\I \, an\, {\mathrm
  Im}\theta\}_{n=1,2,\ldots}$. We now analyze the behaviour of the
eigenvalues of $L_0(\theta)$ under the perturbation $\lambda V(\theta)$. The
strategy is to show that the eigenvalues $\pm E$ are unstable under the
perturbation, and that the degeneracy of the eigenvalue zero is lifted. Note
that the kernel of $L_\lambda$ is non-empty by construction, see
(\ref{eq:kernel}).

{\it Proof of Theorem \ref{specthm}.} 
The central part of the proof is the control of the resonances bifurcating out of the
eigenvalues $\pm E$ and $0$, see Lemma \ref{lemmam1}. A standard
analyticity argument then implies Theorem \ref{specthm}. The latter can be
summarized as follows: one checks that for all complex $z$ with
${\mathrm{Im}z>0}$, $\theta\mapsto \langle
\psi_{\overline\theta},(L_\lambda(\theta)-z)^{-1}\phi_\theta\rangle$ is
analytic in $0<|\theta|<\theta_0$, $\mathrm{Im}\theta>0$, and continuous as
$\mathrm{Im}\theta\downarrow 0$, for a dense set of deformation analytic
vectors $\psi$, $\phi$ (take e.g. finite-particle vectors of Fock space built
from test functions $f(\varkappa)$ with compact support). As is well known, the real
eigenvalues of $L_\lambda(\theta)$ coincide with those of $L_\lambda$, and
away from eigenvalues the spectrum of $L_\lambda$ is purely absolutely continuous.

We now present in more detail the resonance theory.
\begin{lemma}
\label{lemmam1}
Let $\theta'={\mathrm Im}\theta>0$. There is a
$\lambda_1$ (independent of $\theta$) s.t. if
$|\lambda|<\lambda_1\min(1,\theta')$ then in the half-plane $\{{\mathrm
  Im}z\geq -\theta'/2\}$ the spectrum of $L_\lambda(\theta)$ consists of four 
eigenvalues $\varepsilon_\pm(\lambda)$ and $\varepsilon_0(\lambda)$ and $0$ only
(which do not depend on $\theta$).

Moerover, we have $\varepsilon_j(\lambda) =e_j-\lambda^2\varepsilon^{(2)}_j+
o(\lambda^2)$, where $j=+,-,0$, and    
\begin{equation}
\varepsilon^{(2)}_0 =\I \, \mathrm{Im}\,\varepsilon^{(2)}_{\pm}= \I \,
(1+\e^{-2\pi \frac Ea})\xi,  
\label{m18}
\end{equation}
whith $\xi$ given in (\ref{m19}).
\end{lemma}
We remark that the Fermi Golden Rule Condition (\ref{fgrc}) asserts that $\xi>0$. 

{\it Proof of Lemma \ref{lemmam1}.}
By an argument of
stability of the spectrum it is not difficult to show that the spectrum in the
indicated half plane consists of four eigenvalues only.

The position (at second order in $\lambda$) is governed by so-called {\it
  level shift operators}, see e.g. \cite{MLSO} and references therein. We
explain this with the help of the {\it Feshbach map} \cite{BFS}.  

Let $e$ be an eigenvalue of $L_0(\theta)$ and denote the corresponding
(orthogonal) eigenprojection by $Q_e=P_e P_\mathrm{vac}$, where $P_e$ is the
spectral projection of $L_{\mathrm D}$ onto $e$ and $P_\mathrm{vac}$ projects
onto the vacuum in $\cal F$. Set $\overline Q_e:=\bbbone- Q_e$ and denote by
$\overline X^e=\overline Q_e X \overline
Q_e\upharpoonright_{\mathrm{Ran}\overline Q}$ the restriction of an operator
$X$ to $\mathrm{Ran}\overline Q$. A standard estimate using Neumann series
shows the following fact.
\begin{lemma}
\label{lemmam3}
There is a constant $\lambda_2$ (independent of $\theta$) s.t. if $|\lambda|<
\lambda_2 \min(E,\theta')$ then, for each eigenvalue $e$ of $L_0$, the open
ball of radius $\theta'/2$ around $e$, $B(e,\theta'/2)$, belongs to the
resolvent set of $\overline L_\lambda^e(\theta)$. 
\end{lemma}
It follows from Lemma \ref{lemmam3} that the {\it Feshbach map}
\begin{equation}
F_{e,z}(L_\lambda(\theta)) := Q_e\left( e -\lambda^2 V(\theta)\overline Q_e
  (\overline L_\lambda^e(\theta)-z)^{-1}\overline Q_e V(\theta)\right) Q_e
\label{m14}
\end{equation}
is well defined for all $z\in B(e,\theta'/2)$. This map has the following
remarkable {\it isospectrality property} \cite{BFS}: for all $z\in
B(e,\theta'/2)$, 
\begin{equation}
z\in\mathrm{spec}(L_\lambda(\theta)) \Longleftrightarrow
z\in\mathrm{spec}\big(F_{e,z}(L_\lambda(\theta))\big).
\label{m15}
\end{equation}
Thus it suffices to examine the spectrum of the operator
$F_{e,z}(L_\lambda(\theta))$ which acts on the finite dimensional space
$\mathrm{Ran}Q_e$. We expand the resolvent in (\ref{m14}) around $\lambda=0$
and consider spectral parameters $z=e+O(\lambda)$ to obtain
$$
F_{e,z}(L_\lambda(\theta)) = Q_e\left( e -\lambda^2 V(\theta) \overline Q_e
  (\overline L_0(\theta)-e)^{-1}\overline Q_e V(\theta)\right) Q_e +o(\lambda^2),
$$
where $\lim_{\lambda\rightarrow 0}o(\lambda^2)/\lambda^2=0$. We now use
analyticity in $\theta$ to conclude that
\begin{equation}
F_{e,z}(L_\lambda(\theta)) = Q_e\left( e -\lambda^2 V \overline Q_e
  (\overline L_0-e -\I 0_+)^{-1}\overline Q_e V\right) Q_e +o(\lambda^2),
\label{m16}
\end{equation}
where $\I 0_+$ stands for the limit of $\I \varepsilon$ as
$\varepsilon\downarrow 0$. The operators 
$$
\Lambda_e:= Q_e V \overline Q_e
  (\overline L_0-e -\I 0_+)^{-1}\overline Q_e VQ_e
$$ 
are called level shift operators. For $e=\pm
  E$ they reduce in the present case simply to numbers
  ($\dim\mathrm{Ran}Q_e=1$), while  $\Lambda_0$
  corresponds here to a $2\times 2$ matrix. Using the expression
  (\ref{m6}) for $V$ one can calculate explicitly the level shift operators (see
  also \cite{BFS,M,MLSO} for more detail on explicit calculations in related models). 
\begin{lemma}
\label{lemmam4}
In the basis $\{|-,-\rangle, |+,+\rangle\}$ of $\mathrm{Ran}Q_0$ we have  
\begin{equation*}
\Lambda_0=\I\xi \e^{-\pi \frac Ea} 
\left[
\begin{array}{cc}
\e^{-\pi \frac Ea} & -1\\
-1 & \e^{\pi \frac Ea}
\end{array}
\right],\ \ \mbox{and}\ \ \mathrm{Im}\Lambda_{\pm E} = (1+\e^{-2\pi \frac
  Ea})\xi\, ,
\end{equation*}
where $\xi$ is given in (\ref{m19})
\end{lemma}
{\it Remark.} The Gibbs state of the detector at inverse temperature
$\beta=2\pi/a$ (represented by a vector $\propto [1, \e^{-\pi E/a}]$) spans
the kernel of $\Lambda_0$.

This lemma together with (\ref{m16}) and the isospectrality (\ref{m15}) shows
the expansions (\ref{m18}) and (\ref{m19}). This proves Lemma
\ref{lemmam1}, and at the same token, concludes the proof of  Theorem \ref{specthm}. \hfill $\blacksquare$

\section{Proofs of Proposition \ref{dynproposition} and of Lemma \ref{lem:standardliouvillean} }
\label{dynpropproofsect}

{\it Proof of Proposition \ref{dynproposition}.\ }The coupled Liouville opertor (\ref{eq:coupledliouvillean}) has the form
$\widetilde{L}_\lambda= L_0+\lambda I$, where $I=\tilde G Q[\tilde g]$ with $\tilde G=A+A^\dagger$ and
$\tilde g(x)=a\delta(x^0)x^1\rho(x_*( x))$ has support in
$W_{\mathrm R}$. Essential selfadjointness of $\widetilde{L}_\lambda$ can
easily be shown using the Glimm--Jaffe--Nelson commutator theorem, see
e.g. \cite{FM}, Section 3.

The Araki-Dyson series expansion gives (weakly on a dense set)
\begin{eqnarray}
\e^{\I t \widetilde{L}_\lambda} M \e^{-\I t \widetilde{L}_\lambda}
 &=& 
\sum_{n=0}^\infty\lambda^n \int_0^t\d t_1\int_{t_1}^t\d
t_2\cdots\int_{t_{n-1}}^t \d t_n 
\Big[ \tilde G(t_1)Q[\tilde g\circ B_{-at_1}], \Big[\cdots\nonumber \\
&& \cdots \Big[\tilde G(t_n)Q[\tilde g\circ B_{-at_n}], \e^{\I tL_0} M \e^{-\I
  tL_0}\Big]\cdots\Big]\Big],
\label{arakidyson}
\end{eqnarray}
where we set $\tilde G(t)=\e^{\I t L_{\mathrm D}}G \e^{-\I t L_{\mathrm D}}$. For $M\in {\cal A}$ any element $M'\in {\cal A}'$ commutes
termwise with the series (\ref{arakidyson}), hence $M'\e^{\I t
  \widetilde{L}_\lambda} M \e^{-\I t \widetilde{L}_\lambda}=\e^{\I t
  \widetilde{L}_\lambda} M \e^{-\I t \widetilde{L}_\lambda} M'$. Therefore we have $\e^{\I t
  \widetilde{L}_\lambda} {\cal A} \e^{-\I t \widetilde{L}_\lambda}\in {\cal
  A}$.
\hfill $\blacksquare$

{\it Proof of Lemma \ref{lem:standardliouvillean}.\ }
Essential selfadjointness is shown using the Glimm--Jaffe--Nelson
commutator theorem, see e.g. \cite{FM} Section 3.1. The fact that $\tilde
L_\lambda$ and $L_\lambda$ define the same dynamics on $\mathcal A$ is easily
derived by using that $L_\lambda-\tilde L_\lambda$ belongs to the commutant
of $\mathcal A$ (and e.g. applying the Trotter product formula), see also
\cite{FM}. \hfill $\blacksquare$


\end{document}